\documentclass[aps,preprint,a4paper,10pt,pre,showpacs,twocolumn, superscriptaddress]{revtex4-1} 

\usepackage{graphicx}
\usepackage{amsmath}
\usepackage{amssymb}
\usepackage{color}

\newcommand{\psib}{\bar{\psi}}
\newcommand{\pert}{\alpha}

\begin{document}

\title{Relaxation of curvature induced elastic stress by the Asaro-Tiller-Grinfeld instability}

\author{C. K\"ohler} 
\affiliation{Institute of Scientific Computing, Technische Universit\"at Dresden, Germany}
\author{R. Backofen}
\affiliation{Institute of Scientific Computing, Technische Universit\"at Dresden, Germany}
\author{A. Voigt}
\affiliation{Institute of Scientific Computing, Technische Universit\"at Dresden, Germany}
\affiliation{cfAED, Technische Universit\"at Dresden, Germany}


\begin{abstract}
A two-dimensional crystal on the surface of a sphere experiences elastic stress due to the incompatibility of the crystal axes and the curvature. A common mechanism to relax elastic stress is the Asaro-Tiller-Grinfeld (ATG) instability. With a combined numerical and analytical approach we demonstrate, that also curvature induced stress in surface crystals can be relaxed by the long wave length ATG instability. The numerical results are obtained using a surface phase-field crystal (PFC) model, from which we determine the characteristic wave numbers of the ATG instability for various surface coverages corresponding to different curvature induced compressions. The results are compared with an analytic expression for the characteristic wave number, obtained from a continuum approach which accounts for hexagonal crystals and intrinsic PFC symmetries. We find our numerical results in accordance with the analytical predictions.
\end{abstract}

\maketitle

\section{Introduction}

In material sciences nano structures are of crucial importance, as they often define the macroscopic properties of the material. The kinetic effects occurring upon the formation of these structures are widely studied, well understood and often used to control the formation process. However, under elastic stress also an interface at rest can develop an instability and lead to the formation of nano structures. This stress-driven instability, known as Asaro-Tiller-Grinfeld (ATG) instability was first studied by Asaro and Tiller \cite{Asaro1972:MetallTrans:3:1789} and later independently by
Grinfeld \cite{Grinfield1986:SovPhysDokl:31:831} and Srolovitz \cite{Srolovitz1989:ActaMetall:37:621}. These authors studied the linear instability of a planar interface of a stressed solid and found that the surface is unstable for perturbations with wave numbers less than a critical value. The instability is manifested by mass transport. The elastic stress in the solid is a destabilizing factor, while the interfacial energy is a stabilizing one and their interplay leads to interface modulations relaxing the elastic stress. It has been extensively studied theoretically and numerically for a wide range of different stress driven rearrangement instabilities, see \textit{e.g.} \cite{Spenceretal_PRL_1991,Gao_JMPS_1994,Kassneretal_EPL_1994,Durand1996:PRL:76:3013,Kuktaetal_JMPS_1997,Zhangetal_JMPS_1999,Savina2003:PRE:67:021606,Raetzetal_JCP_2006}. More recently, the connection between the original continuum formulation and a crystal of discrete constituents was successfully established  \cite{Wu2009:PRB:80:125408,Huang2008:PRL:101:158701,Huang2010:PRB:81:165421,Tegze2009:PRL:103:035702}.

Less understood is the role of elastic stress, which arises from curvature effects for crystals on curved surfaces. Such a situation can be found \textit{e.g.} in the cases of coatings, the assembly of biomembranes, the formation of molecular monolayers or in the packing of filament bundles \cite{Roos2010:NatPhys:6:733,Bruss2012:PNAS:109:10781,DeVries2007:Science:315:358}. The natural lattice packing of these two dimensional crystals is incompatible with the curvature of the surface, since the symmetry axes of the crystals are bent by the curvature, leading to stressed crystals. The influence of this curvature induced elastic stress and its relaxation is under investigation in this letter. In \cite{Mengetal_Science_2014} an elastic instability of a growing colloidal crystal is considered experimentally on a spherical droplet and the behavior is analyzed using a continuum theory. Here, we will instead consider a two-dimensional crystal on a spherical surface at rest and  account for discrete constituents of the crystal by using the phase-field crystal (PFC) approach introduced in \cite{Elder2002:PRL:88:245701, Elder2004:PRE:70:051605}, see also the review \cite{Emmerich2012:AiP:61:665} for the wide applicability of the modeling approach in hard and soft matter systems. The PFC model was also successfully applied to crystals on curved surfaces \cite{Backofen2010:PRE:81:025701, Schmid2014:SM:10:4694, Alandetal_PF_2011, Alandetal_PRE_2012}, but mainly focusing on defects describing grain boundary scars \cite{Bauschetal_Science_2003} and pleats \cite{Irvineetal_Nature_2010}, or properties of Pickering emulsions and Bijels \cite{Stratfordetal_Science_2005}. A comprehensive investigation of curvature induced stress relaxation using the PFC model is missing and will be provided in this letter.

We will briefly introduce the PFC model and the numerical approach and numerically investigate the relaxation of curvature induced stress. The results are compared with an analytical continuum model taking into account the hexagonal structure of the crystal and the intrinsic symmetry of the PFC approach. The comparison allows us to identify the relaxation as an ATG instability.

\begin{figure}[t]
\center
	\includegraphics[width=0.13\textwidth]{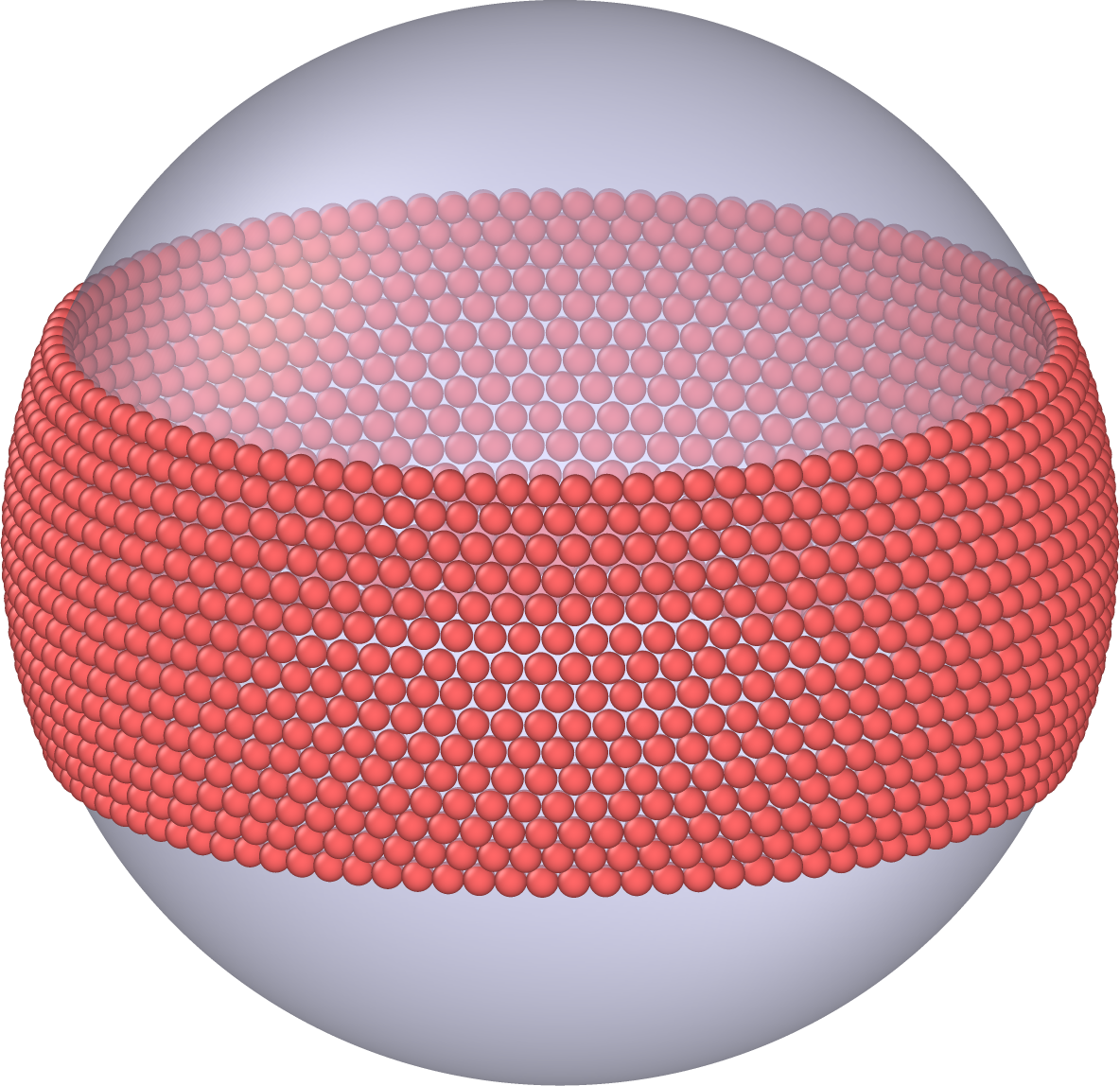}\hskip 0.5cm
	\includegraphics[width=0.13\textwidth]{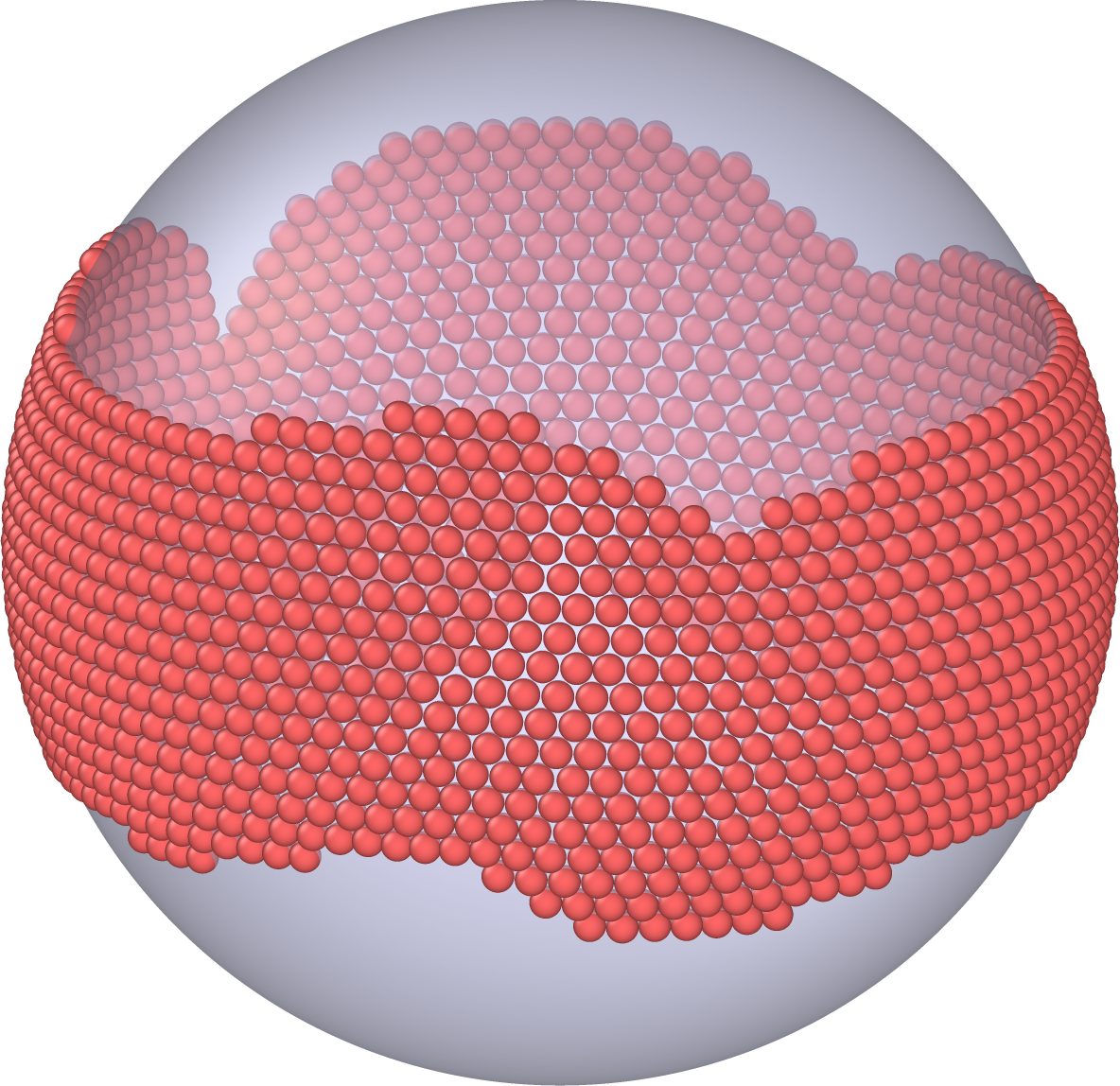}\hskip 0.5cm
	\includegraphics[width=0.13\textwidth]{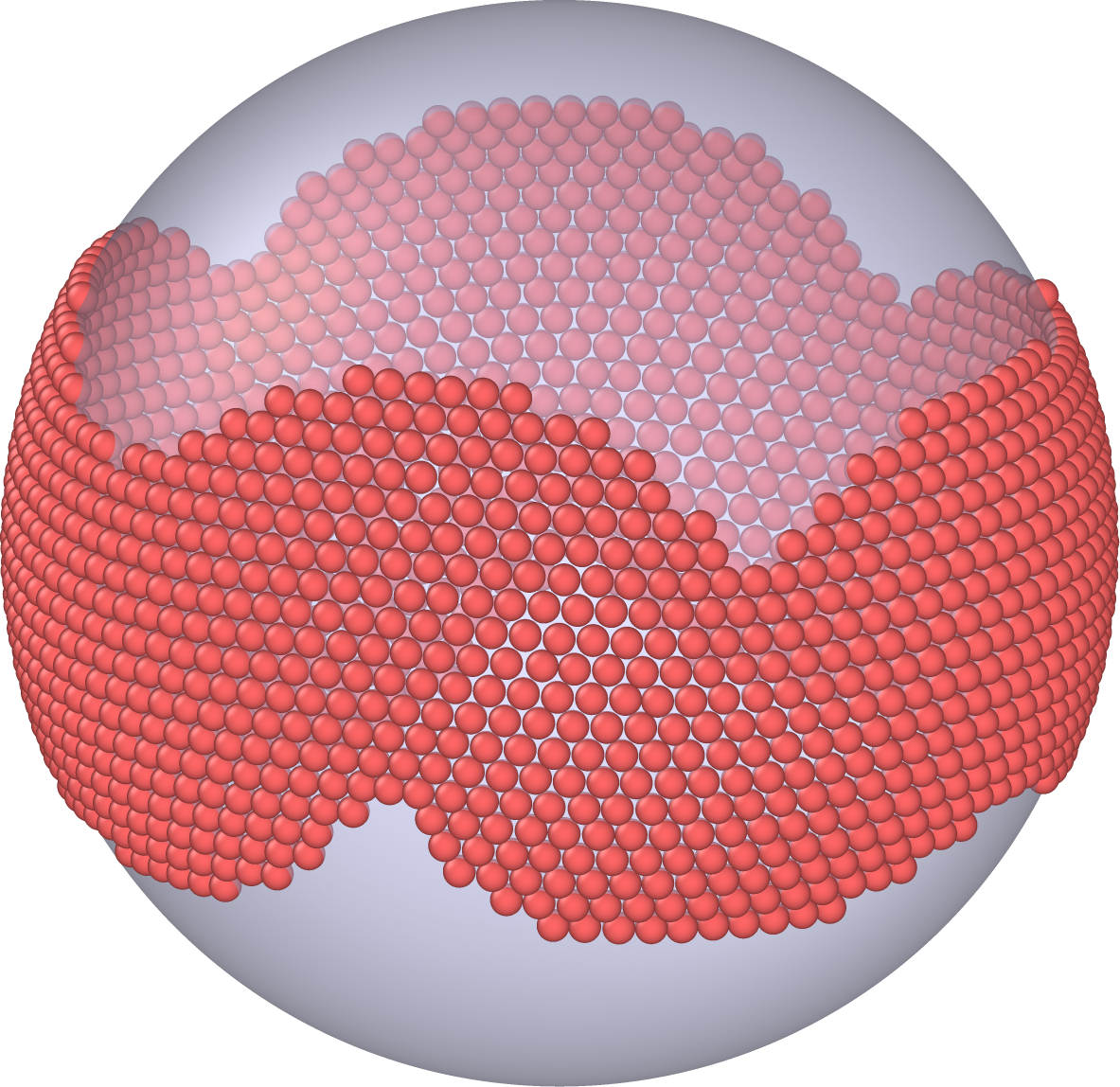}
	\caption{The temporal evolution (left to right) of a ribbon of stressed atoms on the surface of a sphere. The curvature induced stress enhances initially small crystal interface perturbations, which grow exponentially in time. For visualization purposes the maxima in the particle density field are extracted and considered as atoms.}
  	\label{fig:ATG_sphere}
\end{figure}


\section{Phase-Field Crystal Model}
The phase-field crystal (PFC) model is described by an energy functional in terms of the reduced particle density $\psi$ \cite{Backofen2010:PRE:81:025701}
\begin{equation}
  \label{eq:pfc_energy}
  F[\psi] = \int_\Gamma - |\nabla_\Gamma \psi|^2 + \frac{1}{2} |\Delta_\Gamma \psi|^2 + f(\psi) \; \mathrm{d} \Gamma,
\end{equation}
where $f(\psi) = \frac{1}{2}(1 - r) \psi^2 + \frac{1}{4} \psi^4$. $\Gamma$ denotes the curved surface and $\nabla_\Gamma$ and $\Delta_\Gamma$ are the corresponding surface gradient and surface Laplacian. Within our numerical consideration $\Gamma$ will be a sphere. The only parameters are $r$, corresponding to an undercooling and the average density of the system $\psib$. Depending on the parameter $r$ the energy functional is minimized by periodic and/or constant solutions, modelling a crystal and its melt, respectively \cite{Elder2004:PRE:70:051605}. The temporal evolution of the system is given by the $H^{-1}$ gradient flow
\begin{equation}
  \partial_t\psi = \Delta_\Gamma \frac{\delta F[\psi]}{\delta\psi}
  \label{eq:pfc:tempevol}
\end{equation}
making $\psi$ a conserved quantity. In \cite{Backofenetal_MMS_2011} a derivation of this equation from a surface dynamic density functional theory (DDFT) is sketched, following the detailed derivation of the PFC equation in flat space in \cite{Elderetal_PRB_2007,Teeffelenetal_PRE_2009}.

Within a one mode approximation in flat space \cite{Elder2004:PRE:70:051605} the periodic solution is given in the $(x,y)$-plain by 
\begin{equation}
  \psi_p=A\left[\cos\left(qx\right)\cos\left(\frac{q}{\sqrt{3}}y\right)-\frac{\cos\left(\frac{2q}{\sqrt{3}}y\right)}{2}\right]+\psib
  \label{eq:pfc:oma_per}
\end{equation}
with equilibrium wave number $q$ and amplitude $A$ defined as
\begin{equation}
  q=\frac{\sqrt{3}}{2}, \qquad A=\frac{4}{5}\left(\psib+\frac{1}{3}\sqrt{-15r-36\psib^2}\right).
\end{equation}
Correspondingly, the equilibrium wave length is ${a_0=2\pi/q}$. We will use this one mode approximation to initially set up a curvature induced stressed configuration.


\section{Numerical simulation}
Now, we numerically investigate the relaxation of such a curvature induced stressed configuration. As in the classical ATG instability, only wavelengths of interface perturbations above some critical value can be expected to grow exponentially. We therefore need large, stressed crystals and start with a one mode approximation of a ribbon wrapped completely around the equator of a sphere with radius $R=100a_0/2\pi$ (see fig.~\ref{fig:ATG_sphere} for an overview). Its initial width is adjusted such that it closely corresponds to the surface coverage as determined by the PFC parameter $\psib$. This allows to keep crystal growth at bay in order to avoid competing dynamic instabilities (\textit{e.g.} Mullins-Sekerka). As $\psib$ steers the width of the crystal ribbon and each new particle layer is exposed to increasing stress due to the curvature of the sphere, the width can be used to realize setups with differently stressed crystals. We choose the PFC parameter $r=-0.25$ to ensure short simulation times and $\psib=-0.32$, $\psib=-0.31$, $\psib=-0.30$ and $\psib=-0.29$ for increasingly stressed crystals. The PFC equations \eqref{eq:pfc:tempevol} are solved by using a basis decomposition into spherical harmonics combined with a semi implicit Euler time discretization \cite{Backofenetal_MMS_2011, Schaeffer2013:GGG:14:751}. In order to shorten simulation times, small amplitude noise was added in each time step.

\begin{figure*}[th]
\center
	\includegraphics[width=0.9\textwidth]{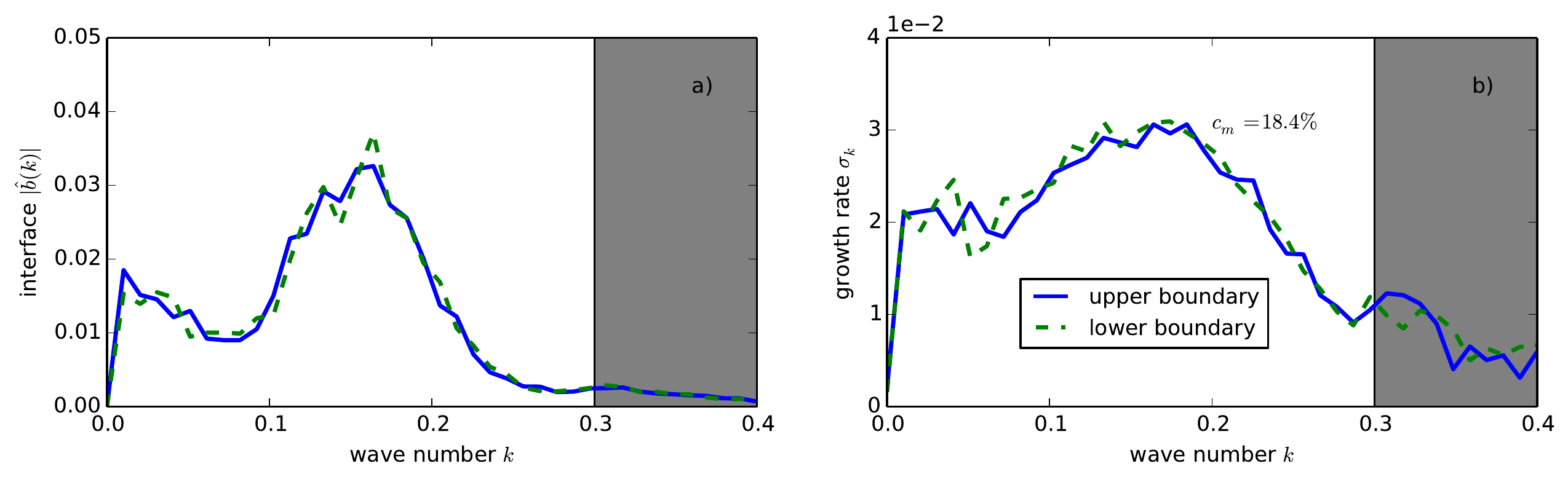}
	\caption{Exemplary interface spectra (a) and growth rates (b) averaged over 20 single noise realizations for a compression $c_m=18.4\%$ ($\psib=-0.29$). The shaded area indicate spectral domains, where the values of the growth rate are non-physical.}
  	\label{fig:gr_100_differentstress}
\end{figure*}

Upon temporal evolution, we determine the upper and lower interface of the crystal, denoted by $b(x)$ where $x$ is a  longitude coordinate. The interface mean values are constant after an initial relaxation of the initial condition and before crystalline defects are incorporated at a late stage of the interface modulation. The interfacial Fourier components $\hat{b}(k)$ are calculated and the amplitude for each wave number $k$ is monitored over time. Subsequently, in order to obtain the growth rate $\sigma_k$ for each wave number $k$, we fit the obtained data to an exponential function $\propto \exp{\left(\sigma_kt\right)}$ using the time interval of constant mean interface.

\begin{figure}
\center
    \includegraphics[width=0.45\textwidth]{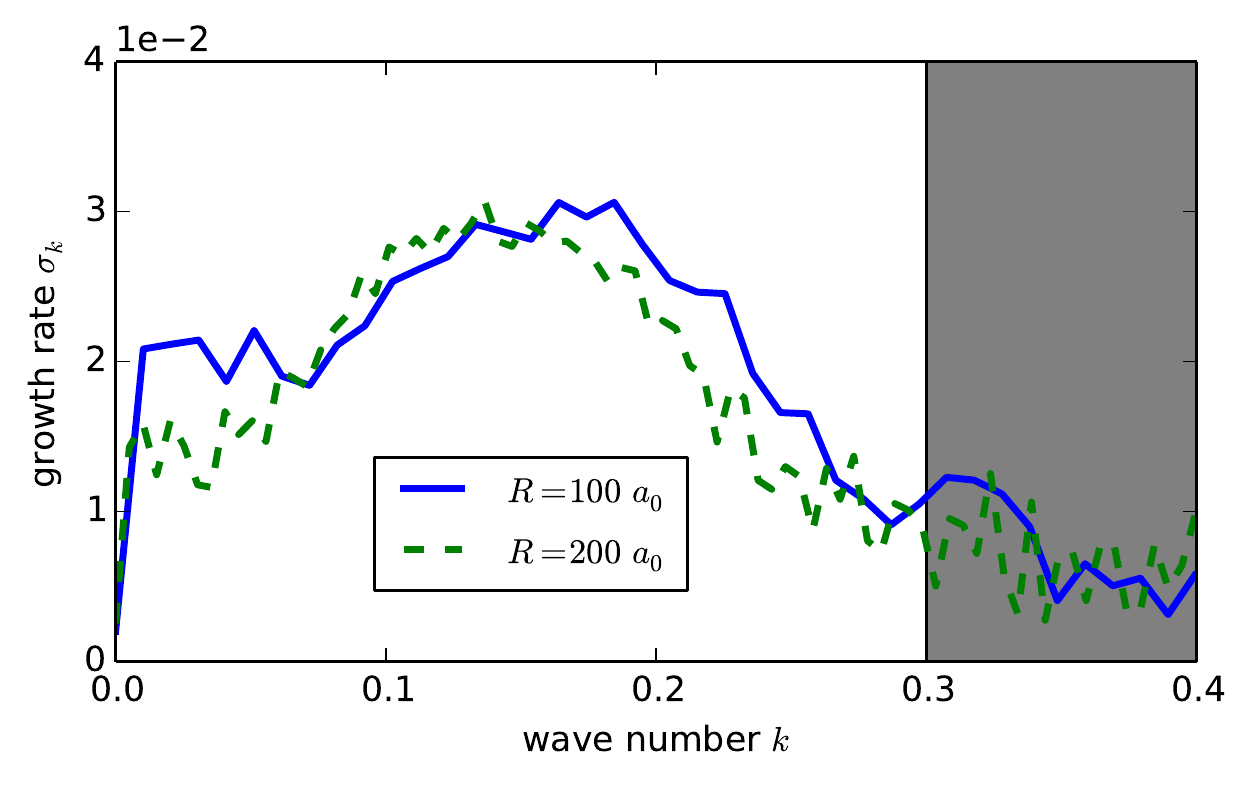}
    \caption{The equality of growth rates for surface radii $R=100~a_0$ and $R=200~a_0$ demonstrates their independence on the number of particle layers ($\psib=-0.29$).}
    \label{fig:ATG_NumCheck}
\end{figure}

We introduce the compression $c_m=(a_0-a)/a_0$ with the actual and equilibrium particle distance at the interface $a$ and $a_0$, respectively. In fig.~\ref{fig:gr_100_differentstress} the numerically obtained growth rate and interface spectrum $\hat{b}(k)$ are exemplarily shown for $\psib=-0.29$. Defining the mean value of the interface as the crystal interface, this corresponds to a compression at the crystal interface of $c_m= 18.4\%$. We extract a maximum growth rate of $\sigma_k=3.0\times 10^{-2}$ for the wave number $k_{max}=0.18$. The slightly noisy structure of the growth rates and spectra in fig.~\ref{fig:gr_100_differentstress} originates from the noise imposed in the numerical simulations. Unfortunately, we also observe considerably large values for the growth rate for wave numbers, where the according spectra of the interface lacks contributions. These non-physical contributions are present only due to the rigorous application of our $\propto e^{\sigma_k t}$ fitting procedure, although spectral components exhibit no exponential behavior at all in this spectral regime. Consequently, these parts of the curves should not be taken into account, as suggested by the shaded areas in fig.~\ref{fig:gr_100_differentstress}. In fig.~\ref{fig:ATG_NumCheck} we demonstrate for the same parameter setting, that the absolute values of the maximum growth rate $\sigma_{k,max}$ and wave number of maximum growth $k_{max}$ do not depend on the number of particle layers necessary to realize a certain coverage of the sphere (and thus a certain curvature induced stress at the crystal interface), as a doubling of the sphere radius (and thus doubling the number of layers) does not considerably alter the previously found values.

Further data of $k_{max}$ for different compression rates $c_m$ appear as solid dots in fig.~\ref{fig:ATG_kmax}. For the additional data $c_m=7.5\%$ ($\psib = -0.32$), $c_m=10.2\%$ ($\psib = -0.31$) and  $c_m=15.3\%$ ($\psib = -0.30$), we observe increasing maximum values of $\sigma_k=1.4\times 10^{-4}$, $\sigma_k=1.8\times 10^{-3}$ and $\sigma_k=1.6\times 10^{-2}$ and increasing wave numbers of maximum growth rate $k_{max}=0.04$, $k_{max}=0.09$ and $k_{max}=0.14$.

These increasing growth rates $\sigma_k$ and wave numbers $k_{max}$ for increasing stress are in accordance with the original continuum ATG theory \cite{Asaro1972:MetallTrans:3:1789, Grinfield1986:SovPhysDokl:31:831, Srolovitz1989:ActaMetall:37:621}, as well as with previous observations from numerical simulations within the amplitude equation approach \cite{Huang2010:PRB:81:165421} in flat space. In  \cite{Huang2010:PRB:81:165421} a 'perfect relaxation' condition was formulated. That condition assumes, that a crystal of discrete constituents reaches a completely stress free state, when it includes a certain number of defects. Equally distributing these defects along the crystal interface defines a wave number. This wave number of maximum stress relaxation is plotted as function of the compression $c_m$ in fig.~\ref{fig:ATG_kmax} and nicely agrees with our numerical results, even if the origin of stress is different. Additionally, we calculate the most unstable wave number $k_{max}$  within a continuum elasticity model. 

\section{Continuum elasticity} 
Using the continuum elasticity theory with hexagonal symmetry and intrinsic PFC symmetries, we derive expressions for the wave number of a maximum growth rate $k_{max}$. We thereby closely follow \cite{Cantat1998:PRE:58:6027}, where the ATG instability for an isotropic continuous medium was analyzed in flat space.

\begin{figure}
\center 
    \includegraphics[width=0.45\textwidth]{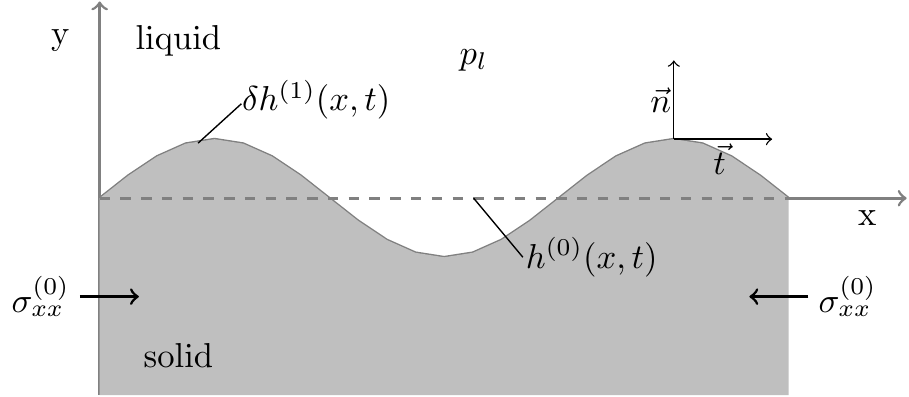}
    \caption{The geometry of the continuum elasticity model. Under external stress $\sigma_{xx}^{(0)}$ modulations $\pert h^{(1)}$ of the unperturbed interface $h^{(0)}$ grow.}
    \label{fig:ATG_sketch}
\end{figure}

We start with the general expression for the elastic energy (see, \textit{e.g.}, \cite{Landau1975})
\begin{equation*}
  F=F_0+\frac{1}{2}C_{ijkl}\epsilon_{ij}\epsilon_{kl},
\end{equation*}
where we use the elastic constants $C_{ijkl}$ and the strain tensors $\epsilon_{ij} = \frac{1}{2}(\partial u_i/\partial x_j + \partial u_j/\partial x_i)$ with the displacement fields $u_i$. The indices obey $i,j,k,l \in \{x,y\}$ for two spatial coordinates $x_1=x$, $x_2=y$, see fig. \ref{fig:ATG_sketch}. We use Voigt's notation $xx \rightarrow 1$, $yy\rightarrow 2$ and $xy=yx\rightarrow 3$. Exploiting the intrinsic symmetries for the elastic constants and strain tensors and additionally accounting for hexagonal crystal and intrinsic PFC symmetry results in
\begin{align*}
  C_{11}&=C_{22}=3C_{33}=3C_{12},\qquad  C_{13}=C_{23}=0.
\end{align*}
Thus, the free energy reads
\begin{align*}
  F=F_0 + &\frac{C_{33}}{2}\left(3\epsilon_{xx}^2 + 3\epsilon_{yy}^2 + 4\epsilon_{xy}^2 + 2\epsilon_{xx}\epsilon_{yy}\right).
\end{align*}
The stress tensor $\sigma_{ij} = \partial F / \partial \epsilon_{ij} = C_{ijkl}\epsilon_{kl}$ is related to the strain tensor via
\begin{equation*}
  \label{eq:stresstensor}
  \sigma_{ij}=4C_{33}\epsilon_{ij} + C_{33} \delta_{ij}\left(-2\epsilon_{ij} + \epsilon_{kk}\right),
\end{equation*} 
and vice versa
\begin{equation*}
  \epsilon_{ij}=\frac{\sigma_{ij}}{4C_{33}} + \frac{1}{4C_{33}}\delta_{ij}\left(\sigma_{ij}-\frac{\sigma_{kk}}{2}\right), 
\end{equation*}
and the equilibrium equation to solve reads
\begin{equation}
  \frac{\partial \sigma_{ik}}{\partial x_k}=0
\end{equation}
which is satisfied for
\begin{equation*}
  \sigma_{xx}=\frac{\partial^2 \chi}{\partial y^2},\quad \sigma_{xy}=-\frac{\partial^2 \chi}{\partial x\partial y},\quad\sigma_{yy}=\frac{\partial^2 \chi}{\partial x^2}
\end{equation*} 
with an arbitrary Airy stress function $\chi=\chi\left(x,y\right)$. At the crystal interface we formulate 
\begin{equation}
  \sigma_{nn}=-p_l,\quad \sigma_{nt}=0,
  \label{eq:equibfront}
\end{equation}
where $p_l$ is the liquid pressure and 
\begin{equation*}
  \sigma_{nn}=n_i\sigma_{ij}n_j, \quad \sigma_{tt}=t_i\sigma_{ij}t_j, \quad \sigma_{nt}=n_i\sigma_{ij}t_j
\end{equation*}
are the normal and tangential components of the stress tensor. Describing the crystal interface as $y=h(x,t)$, the normal and tangent vectors of the interface are given by
\begin{align*}
  (t_x,t_y)&=(1,\partial h / \partial x)/\sqrt{1+(\partial h / \partial x)^2},\\
  (n_x,n_y)&=(-\partial h / \partial x,1)/\sqrt{1+(\partial h / \partial x)^2}.
\end{align*} 

We now solve eq.~\eqref{eq:equibfront} in a perturbative manner. We assume $\pert$ to be a small parameter, which describes the strength of the interface modulation. The interface $h(x,t)$ now reads:
\begin{align}
  h = h^{(0)} + \pert h^{(1)} + \pert^2 h^{(2)} + \cdots.
  \label{eq:interface_perturb}
\end{align}
Similarly, also the remaining variables are expanded in a power series in $\pert$
\begin{align*}
  \epsilon_{ij}=&\epsilon_{ij}^{(0)}+\pert \epsilon_{ij}^{(1)} + \pert^2 \epsilon_{ij}^{(2)} + \cdots,\\  
  \sigma_{ij}=&\sigma_{ij}^{(0)} + \pert\sigma_{ij}^{(1)} + \pert^2\sigma_{ij}^{(2)} + \cdots,\\
  \chi_{ij}=&\chi_{ij}^{(0)}+\pert\chi_{ij}^{(1)} + \pert^2\chi_{ij}^{(2)} + \cdots.
\end{align*}
Plugging in the expansions and reordering all terms by powers of $\pert$, eq.~\eqref{eq:equibfront} reads up to first order
\begin{align}
  \label{eq:equib_nn_pert}
  0 =& p_l + \sigma_{yy}^{(0)} + \pert\left(\sigma_{yy}^{(1)}-2{h^{(1)'}}\sigma_{xy}^{(0)}\right)\\
  \label{eq:equib_nt_pert}   
  0=& \sigma_{xy}^{(0)} + \pert\left(- \sigma^{{(0)}}_{{xx}} {h^{(1)'}} + \sigma^{{(1)}}_{{xy}} + \sigma^{{(0)}}_{{yy}} {h^{(1)'}}\right),
\end{align} 
where $'$ denotes the derivative w.r.t $x$. Evaluating these equations to zeroth order in $\pert$ gives
\begin{equation}
  \sigma_{yy}^{(0)} =-p_l, \quad \sigma_{xy}^{(0)}=0
  \label{eq:stress_zeroth_order}
\end{equation} 
and $\sigma_{xx}^{(0)}$ is the applied stress. To proceed with the first order in $\pert$, we make the ansatz
\begin{equation}
  \chi^{(1)} = \left(A + By\right)\exp\left(i kx + ky + \omega t \right)
\end{equation}
and determine the constants $A$, $B$ by evaluating the interface conditions \eqref{eq:equib_nn_pert} and \eqref{eq:equib_nt_pert} to first order in $\pert$. Assuming further, that the first order perturbation of the flat ($h^{(0)}(x)=const$) interface obeys
\begin{equation}
  h^{(1)} = h_{11} \exp\left(i kx + \omega t \right),
  \label{eq:surface_mode_first_order}
\end{equation}
we end up with
\begin{align}
  \sigma_{xx}^{(1)} = & -2 k \sigma_0  h_{11} \exp\left(i k x + \omega t\right),\nonumber\\
  \sigma_{xy}^{(1)} = & i k \sigma_0  h_{11} \exp\left(i k x + \omega t\right),\nonumber\\
  \sigma_{yy}^{(1)} = & 0,
  \label{eq:stress_first_order}
\end{align}
where we introduced $\sigma_0=\sigma_{xx}^{(0)}-\sigma_{yy}^{(0)}$.

The temporal evolution of the surface perturbation $h^{(1)}$ is induced by the solidification of liquid at the crystal interface. The solidification is driven by the difference of the chemical potential between the liquid and solid phase $\Delta{\mu}=\mu_{liquid}-\mu_{solid}$:
\begin{equation}
  \frac{\partial h}{\partial t} = \frac{\partial h^{(0)}}{\partial t} + \pert\frac{\partial h^{(1)}}{\partial t} = \frac{\partial h^{(0)}}{\partial t} + \pert\omega h^{(1)} = f \Delta{\mu},
  \label{eq:perth_mu}
\end{equation}
with some proportionality constant $f$. We encounter the same thermodynamic situation at the crystal interface that is described in detail in \cite{Cantat1998:PRE:58:6027}, appendix A. However, we consider hexagonal crystals, making the mathematical expressions slightly more extensive.

We start with the two phases at equilibrium. When transforming a small mass element at the interface of volume $\delta V$ from liquid into solid, the change of the Gibbs free energy is
\begin{equation}
  \Delta G = \Delta F + \Delta(p_l \delta V) = \delta V \Delta \mu,
  \label{eq:Gibbs_Durham}
\end{equation}  
where $\Delta F$ is the Helmholtz free energy change.  The Helmholtz free energy $\Delta F = \Delta F_i + \Delta F_m$ is composed of the free energy change of the transformed mass element $\Delta F_m$ and an interface contribution $\Delta F_i$. The contribution $\Delta F_i= \gamma \kappa\delta V$ accounts for the change of the interface free energy caused by the interface tension $\gamma$ and the interface curvature $\kappa$. For the interface eq.~\eqref{eq:interface_perturb} the curvature $\kappa$ is given to first order $\pert$
\begin{align*}
  \kappa = &\frac{h''}{\left(1+h'^2\right)^{3/2}} \simeq -k^2\pert h^{(1)},
  \label{eq:curvature_first_order} 
\end{align*}
giving 
\begin{equation}
  \Delta F_i = -\gamma k^2\pert h^{(1)}\delta V.
  \label{eq:dmu_interface}
\end{equation}
The free energy change of the transformed mass element $\Delta F_m$ is given by the work necessary to increase its internal strain to the value of the surrounding solid. The work for an infinitesimal change in strain is
\begin{equation}
  \mathrm{d} f = \sigma_{ij}\mathrm{d}\epsilon_{ij}.
\end{equation}
Together with the change of the volume element $\textrm{d}\left(\delta V\right)$ due to increasing strain, the infinitesimal elastic contribution to $\Delta G$ reads
\begin{align}
  dG_{el} = &\sigma_{ij}d\epsilon_{ij}\delta V+p_ld(\delta V)\nonumber\\
  = &\left[\left(\sigma_{nn}d\epsilon_{nn}+\sigma_{tt}d\epsilon_{tt}\right)-\sigma_{nn}\left(d\epsilon_{nn}+d\epsilon_{tt}\right)\right]\delta V\nonumber\\
  = &\left(\sigma_{tt}-\sigma_{nn}\right)d\epsilon_{tt}\delta V.
  \label{eq:infinitesimal_G_el}
\end{align}
We used the mechanical equilibrium eq.~\eqref{eq:equibfront} by substituting $-p_l=\sigma_{nn}$ and exploiting $\sigma_{nt}=0$, \textit{i.e.} the stress tensor is diagonal in the coordinate system defined by the vectors normal and tangential to the interface. We express $\mathrm{d}\epsilon_{tt}$ and $\left(\sigma_{tt}-\sigma_{nn}\right)$ in terms of the small parameter $\pert$ and discard all terms of higher order than one:
\begin{align}
  \mathrm{d}\epsilon_{tt}=&\frac{\mathrm{d}\sigma_{tt}}{4C_{33}} + \frac{\mathrm{d}\sigma_{xx}^{(0)}}{8C_{33}}-\frac{\mathrm{d}\sigma_{yy}^{(0)}}{8C_{33}} + \pert\frac{\mathrm{d}\sigma_{xx}^{(1)}}{8C_{33}}-\pert \frac{\mathrm{d}\sigma_{yy}^{(1)}}{8C_{33}}\nonumber
\end{align}
and
\begin{align}
  \sigma_{tt}-\sigma_{nn}=& \sigma_0\left(1-2k\pert h^{(1)}(x,t)\right).     
\end{align}
Using the relations \eqref{eq:stress_first_order} for $\textrm{d}\sigma_{xx}^{(1)}$, $\textrm{d}\sigma_{yy}^{(1)}$ and integrating eq.~\eqref{eq:infinitesimal_G_el} over stress values up to the interface values given in eqs.~\eqref{eq:stress_zeroth_order}, \eqref{eq:stress_first_order} results in 
\begin{align*}
   \Delta G_{el} = &\Big[\frac{\sigma_0^2}{8C_{33}}\left(1-4k\pert h^{(1)}\right)+\frac{\sigma_0^2}{16C_{33}}\left(1-2k\pert h^{(1)}\right)\nonumber\\
    &+\frac{\sigma_0^2}{8C_{33}}\left(-k\pert h^{(1)}\right)\Big]\delta V.
\end{align*}
After plugging this together with eq.~\eqref{eq:dmu_interface} into eq.~\eqref{eq:Gibbs_Durham} and dividing by $\delta V$, we arrive at
\begin{align}
  \Delta{\mu} =& \frac{\sigma_0^2}{8C_{33}}\left(1-4k\pert h^{(1)}\right)+\frac{\sigma_0^2}{16C_{33}}\left(1-2k\pert h^{(1)}\right)\nonumber\\
    &+\frac{\sigma_0^2}{8C_{33}}\left(-k\pert h^{(1)}\right)-\gamma k^2\pert h^{(1)},
  \label{eq:dmu_final}
\end{align}
Comparing to eq.~\eqref{eq:perth_mu} we deduce the exponential growth rate $\omega$ of the interface perturbation
\begin{align}
  \omega =& f\left(\frac{3\sigma_0^2}{4C_{33}}k-\gamma k^2\right),
\end{align}
which is maximal for the wave number
\begin{align}
  k_{max} = \frac{3}{8}\frac{\sigma_0^2}{\gamma C_{33}}.
  \label{eq:kmax}
\end{align}
Now, the result for the wave number of maximum growth is adapted to the crystal on the curved surface of a sphere.


\section{Discussion}

\begin{figure}[t]
	\includegraphics[width=0.48\textwidth]{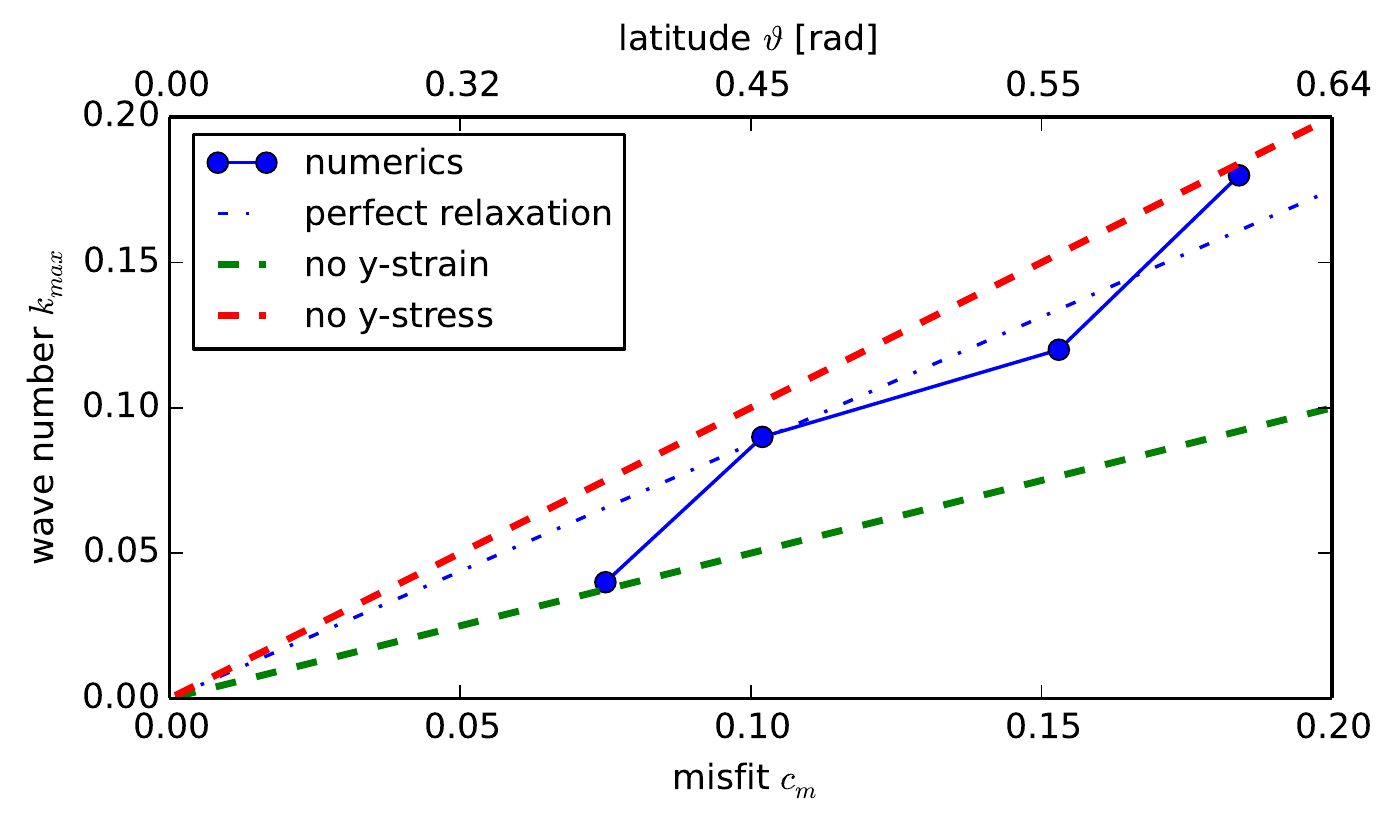}
	\caption{The wave number $k_{max}$ of the maximum growth rate is plotted as a function of the compression $c_m$. For the crystals on the sphere, the compression is curvature induced and related to the latitude $\vartheta$ of the crystal interface $c_m=1-\cos(\vartheta)$. The results from numerical simulations (solid blue line with dots) are bounded by the two limiting cases of zero stress (red dashed line) and zero strain (green dashed line) in y-direction. The blue dash dotted line corresponds to the 'perfect relaxation' condition from \cite{Huang2010:PRB:81:165421} and agrees very well with our results.}
  	\label{fig:ATG_kmax}
\end{figure}

In the derivation of eq.~\eqref{eq:kmax}, all defining quantities were evaluated at the crystal interface. In particular, only the elastic properties and the chemical potential difference of the interface determine the value of $k_{max}$. Thus, we identify the crystal interface from the previous calculations in flat geometry with the interface of the crystal on the curved surface of the sphere.
Because the elastic constant $C_{33}$ can be obtained directly from the PFC parameters $r$ and $\bar{\psi}$ \cite{Elder2004:PRE:70:051605}, the remaining task is to determine the interface tension $\gamma$ and the externally applied stress $\sigma_0$ with regard to the curvature of the spherical surface. We start with the strain in x-direction (parallel to the unperturbed interface $h^{(0)}$) for the crystal on the sphere. The strain is $\epsilon_{xx}^{(0)}=1-\cos(\vartheta)$ for a crystal interface at latitude $\vartheta$, provided the crystal is stress free at the equator ($\vartheta=0$). This emphasizes the difference to flat geometry: the curvature induced compression $c_m=1-\cos(\vartheta)$ increases for larger latitudes, whereas the compression is constant in the flat geometry. For the y-direction (perpendicular to $h^{(0)}$), we can either assume zero strain ($\epsilon_{yy}^{(0)}=0$) or zero stress ($\sigma_{yy}^{(0)}=0$). The two cases correspond to $p_l=-c_m C_{33}$ or to zero liquid pressure $p_l=0$, respectively. The interface tension $\gamma$ is obtained as the ratio of the energy $\mathrm{d}F$ needed to prolong the interface by a certain length $\mathrm{d}L$ and $\mathrm{d}L$, \textit{i.e.} $\gamma=\mathrm{d}F/\mathrm{d}L$. We obtain
\begin{align}
  \sigma_0=&\gamma=\frac{8}{3}C_{33}c_m \qquad \mbox{zero stress},\\
  \sigma_0=&\frac{2}{3}\gamma=2C_{33}c_m \qquad \mbox{zero strain}.
\end{align}
Accordingly, we finally get
\begin{align}
  k_{max}=&c_m \qquad \mbox{zero stress},\\
  k_{max}=&\frac{1}{2}c_m\qquad \mbox{zero strain}.
\end{align}
These two lines are also plotted in fig. \ref{fig:ATG_kmax}. The actual liquid pressure lies between the two limiting cases of no liquid pressure $p_l=0$ and the case where the liquid pressure is so strong that it ensures zero displacement in y-direction $p_l=-c_mC_{33}$. Our numerical results are in between these two limiting lines and thus allows identifying the observed elastic instability as a curvature induced ATG instability.


\section{Summary} 
In summary, we numerically simulated the relaxation of curvature induced elastic stress for crystals on a spherical surface within a PFC model. In agreement with an analytical continuum model accounting for hexagonal crystal and inherent PFC symmetries, the relaxation is mediated by the ATG instability. Accordingly, we found that the elastic stress at the crystal interface is defining the growth rates and the characteristic wavelength of maximum growth of the interface modulations. This situation is different to the elastic instability discussed in \cite{Mengetal_Science_2014}, where the growth of a crystal under the influence of curvature induced stress leads to anisotropic growth. 

Even if the used numerical approach is restricted to the geometry of a sphere using one of the other numerical approaches discussed in \cite{Backofenetal_MMS_2011} any curved surface can be considered or even surface modulations, possibly induced by the stressed crystal \cite{Alandetal_MMS_2012} or resulting from external forces.


\acknowledgments
We acknowledge financial support from DFG through SPP 1296 within project Vo899-7 and computing resources provided by JSC within project HDR06.



\bibliography{library}

\end{document}